\documentclass[aps,prx,twocolumn,final,letterpaper]{revtex4}

\usepackage{appendix}
\usepackage{graphicx}   
\usepackage{import}                         
\usepackage{epstopdf}
\usepackage{amsmath} 
\usepackage{bm}
\usepackage{amssymb}
\usepackage{quotes}
\usepackage{transparent}
\usepackage{dcolumn}
\usepackage{multirow}
\usepackage{cancel} 
\usepackage{mdframed}
\usepackage{color}
\usepackage{bm}
\usepackage{dsfont}
\usepackage{slashed}
\usepackage{enumitem}
\usepackage{braket}

\begin{document}
\rmfamily

\title{Quantum sensitivity of parametric oscillators}


\author{Alex Gu$^{1}$}
\email{alexgu@mit.edu}
\author{Jamison~Sloan$^{1}$}
\author{Charles Roques-Carmes$^{1,2}$}
\email{chrc@stanford.edu}
\author{Seou Choi$^{1}$}
\author{Eric I. Rosenthal$^{2}$}
\author{Michael Horodynski$^{3}$}
\author{Yannick Salamin$^{1,3,4}$}
\author{Jelena Vu\v{c}kovi\'{c}$^{2}$}
\author{Marin Solja\v{c}i\'{c}$^{1,3}$}

\affiliation{$^{1}$ Research Laboratory of Electronics, Massachusetts Institute of Technology, Cambridge, MA 02139, USA\looseness=-1}
\affiliation{$^{2}$ E. L. Ginzton Laboratory, Stanford University, Stanford, CA 94305, USA\looseness=-1}
\affiliation{$^{3}$ Department of Physics, Massachusetts Institute of Technology, Cambridge, MA 02139, USA\looseness=-1}
\affiliation{$^4$ CREOL, The College of Optics and Photonics, University of Central Florida, Orlando, Florida 32816, USA.}

\begin{abstract} 
Many quantum systems exhibit high sensitivity to their initial conditions, where microscopic quantum fluctuations can significantly influence macroscopic observables. Understanding how quantum states may influence the behavior of nonlinear dynamic systems may open new avenues in controlling light-matter interactions. To explore this issue, we analyze the sensitivity of a fundamental quantum optical process -- parametric oscillation -- to quantum initializations. Focusing on optical parametric oscillators (OPOs), we demonstrate that the quantum statistics of arbitrary initial states are imprinted in the early-stage dynamics and can persist in the steady-state probabilities. We derive the ``quantum sensitivity'' of parametric oscillators, linking the initial quantum state to the system's steady-state outcomes, highlighting how losses and parametric gain govern the system's quantum sensitivity. Moreover, we show that these findings extend beyond OPOs to a broader class of nonlinear systems, including Josephson junction based superconducting circuits. Our work opens the way to a new class of experiments that can test the sensitivity of macroscopic systems to quantum initial conditions and offers a pathway for controlling systems with quantum degrees of freedom.
\end{abstract}

\maketitle

\section*{Introduction}

A hallmark feature of nonlinear dynamical systems is that they can exhibit high sensitivity to initial conditions. This can lead to multiple stable solutions, instabilities, and chaotic behavior~\cite{haken2012advanced}. In the quantum domain, interacting systems become even more complex due to an expanded configuration space and the influence of quantum zero-point fluctuations~\cite{arecchi2012instabilities, arecchi1985chaos}. This enhanced sensitivity is also foundational for quantum-enhanced sensing protocols~\cite{zanardi2008quantum, di2023critical, petrovnin2024microwave, alushi2024optimality, beaulieu2024criticality, guo2024quantum}. Quantum fluctuations can destabilize systems that are classically stable, underpinning phenomena such as superradiance in ensembles of emitters~\cite{gross1982superradiance}, the operating phases of lasers and optical parametric oscillators~\cite{siegman1986lasers}, noise seeding in highly nonlinear light sources like supercontinuum sources and free electron lasers~\cite{pellegrini2016physics}. Understanding how nonlinear systems respond to quantum fluctuations and controlling these behaviors with quantum states of light remains a key challenge. Recent studies have begun exploring modifications in light-matter interactions under quantum driving states, including enhanced interactions in cavity quantum electrodynamics with squeezed light~\cite{leroux2018enhancing} and the use of bright squeezed states for high-harmonic generation~\cite{gorlach2023high}, photoelectron emission~\cite{heimerl2024multiphoton}, ionization~\cite{fang2023strong}, and electron-photon interactions~\cite{khalaf2023compton}. There remain, however, large classes of systems whose responses to arbitrary quantum states are yet to be explored.





One such example is parametric amplification, which is a fundamental process that enables interactions among photons or other bosons. 
Parametric amplification and oscillation play a crucial role in generating quantum states of light such as squeezed states, generating and amplifying light at tunable frequencies, and amplifying signals for measurement~\cite{wu1987squeezed, shih1988new, nehra2022few, shaked2018lifting}. Consequently, studying the response of driven-dissipative systems like parametric oscillators to arbitrary quantum inputs is highly relevant. This is especially pertinent given recent advances in integrated platforms that combine multiple quantum functionalities into compact form factors, including single-photon emitters, squeezed state generation, and quantum state measurement~\cite{zhu2021integrated, wang2018integrated, holzgrafe2020cavity, nehra2022few, williams2024ultrashort, lukin20204h, lukin2020integrated,guidry2022quantum, tziperman2024parametric}.

Here, we present a theory of parametric oscillation under arbitrary quantum initial conditions. Above threshold, such systems can evolve into one of two bistable steady states \cite{roques-carmes2023biasing}. We focus first on optical parametric amplification, and develop an analytical theory which shows how the probabilities of these two classical steady-state outcomes are determined by the quantum initial conditions. In particular, our theory explains how the parametric gain of early time evolution can imprint the statistics of some initial quantum state onto the steady-state statistics of classical outcomes. Later, we explain how these results are not restricted to OPOs, and can also apply to other systems, providing an example based on superconducting circuits with a Josephson function parametric amplifier. Our work represents a prototypical example of how driven-dissipative systems can respond to quantum inputs, enabling potential future directions in the control, measurement, and sensing of nonlinear quantum systems.

\begin{figure*}
\centering
\vspace{-0.2cm}
  \includegraphics[scale=0.95]{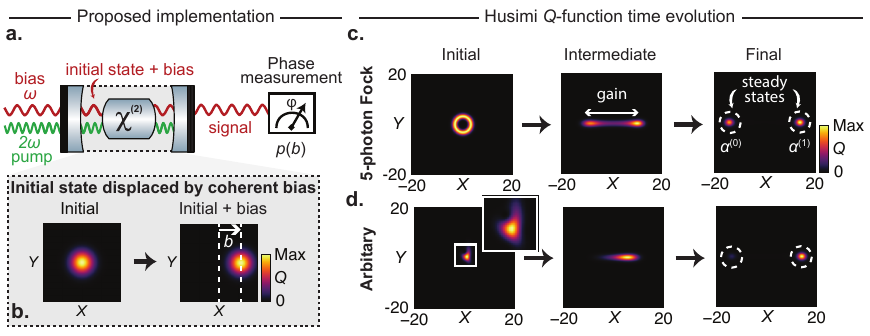}
    \caption{\small \textbf{Bistable driven-dissipative dynamics with arbitrary quantum initial state.} \textbf{a.} Schematic of the proposed setup. Pumping at $2\omega$ induces oscillations at signal frequency $\omega$ via parametric amplification, in a cavity resonant at $\omega$ with a nonlinear $\chi^{(2)}$ crystal, leading to a bistable steady state distribution. The phase of the signal mode is then measured. \textbf{b.} The injected coherent bias $b$ displaces the initial quantum state. \textbf{c, d.} Examples of dynamics with different initial conditions with $5$-photon Fock state (\textbf{c.}) and some arbitrary initial quantum bias $\ket{0,i/2} + \ket{2+i/2,1}$, where $\ket{\alpha,\xi}$ denotes a squeezed coherent state with displacement $\alpha$ and squeeze parameter $\xi$ (\textbf{d.}). In \textbf{c.} we begin with a symmetric Fock state with a small applied bias $b=0.5$. In intermediate stages of time-evolution, gain induces a bifurcation, leading to a steady-state bistable distribution with two Gaussian components. The final steady state probabilities are uneven due to the applied bias. In \textbf{d.} we start with an asymmetrical initial state, which gives a strongly uneven final steady state distribution, even in the absence of an additional bias field $b=0$.}
    \label{fig:1}
    \vspace{-0.3cm}
\end{figure*}


\section*{Results}
\subsection*{Optical parametric oscillators with arbitrary quantum initializations}

In the following, we consider a degenerate biased optical parametric oscillator (OPO) system, consisting of a second-order nonlinear crystal in an optical cavity (Fig.~\ref{fig:1}(a)). Pumping at frequency $2\omega$ induces oscillations at the signal frequency $\omega$ (with corresponding annihilation operator $a$) via parametric amplification, leading to two steady states with phases $\alpha^{(1)} = 0$ and $\alpha^{(0)}=\pi$ when above threshold. Additionally, a coherent bias field of amplitude $b$ is injected into the cavity at frequency $\omega$. In this case, the quantum dynamics of the system are governed by the following Heisenberg-Langevin equation:
\begin{equation}\dot{a} = -a + \lambda a^\dagger - g^2(a^\dagger a)a + F(t) + b.\label{eq:1}\end{equation}
Here, $a$ is the annihilation operator of the cavity mode, $\lambda$ is the fraction above threshold (equal to the gain rate divided by cavity decay rate), $g$ is the quantum noise level ($g^2$ is the nonlinearity strength divided by cavity decay rate), and $F(t)$ is an operator-valued Langevin noise.

We then consider the behavior of some arbitrary quantum state initialized in the cavity at $t=0$. The initial state evolves into one of the degenerate steady states with probabilities $p(\alpha^{(1)})$ and $p(\alpha^{(0)})=1-p(\alpha^{(1)})$ which depend on the applied bias and the initial quantum state. For example, with no bias (i.e. vacuum), the two degenerate energy minima are equally likely $p(\alpha^{(0)})= p(\alpha^{(1)}) = 1/2$. However, different behaviors are possible in the presence of more complex initial states. To see this, we show the dynamics of the OPO with two different initial conditions, as visualized through the Husimi $Q$-function (Fig.~\ref{fig:1}(c,d)).
After the state is initialized, (left panel) parametric gain amplifies the state along the $X$-quadrature creating a separation into two lobes (middle panel). Then, nonlinear saturation causes the field to settle into one of the two possible steady-states $\alpha^{(1)}$ and $\alpha^{(0)}$ (right panel). For the Fock state we considered, we added a small bias to the initial state, which leads to the asymmetry in the final steady-state distribution. In both cases, the asymmetry in the initial state is reflected in the final weighting of the two steady states. 



Now, we develop an analytical theory which predicts the bistable probabilities for arbitrary quantum initial conditions and coherent bias fields. Our approach makes use of the insight that, in the low quantum noise regime which is well respected for optical systems with macroscopic photon numbers ($g^2 \ll \lambda - 1$), the steady-state probabilities are influenced almost entirely by the early-time dynamics before nonlinear saturation sets in. 
Under this approximation, the steady-state probabilities are independent of $g$, allowing us to consider only the dynamics of parametric amplification which occur at early time. 
In this case, we can use standard approaches \cite{walls2007quantum} to obtain decoupled stochastic differential equations for the quadrature variables associated with some chosen representation of the density matrix (e.g., Husimi $Q$ function, Wigner, or Glauber-Sudarshan $P$):
\begin{equation}\begin{aligned}
    &\dot{X} = (\lambda-1)X + \sqrt{2}b + \sqrt{1 + \xi(1 - \lambda)} \eta_1(t) \\
    &\dot{Y} = -(\lambda+1)Y + \sqrt{1 + \xi(1+\lambda)} \eta_2(t).\label{eq:sde-maintext}
\end{aligned}\end{equation}
Here, $X,Y$ are the standard quadratures (related to the coherent-state representations via $\alpha = \frac{1}{\sqrt{2}}(X+ iY)$), $\eta_1(t),\eta_2(t)$ are independent standard normal distributions with zero mean and covariance $\langle \eta_i(t)\eta_j(t)\rangle = \delta_{ij}$, and $\xi = 1,0,-1$ for the Husimi $Q$, Wigner, and Glauber-Sudarshan $P$ representations, respectively. Details on the derivation can be found in the Supplementary Information (SI), Section S1. 

\begin{figure}
\centering
\vspace{-0.2cm}
  \includegraphics[scale=0.9]{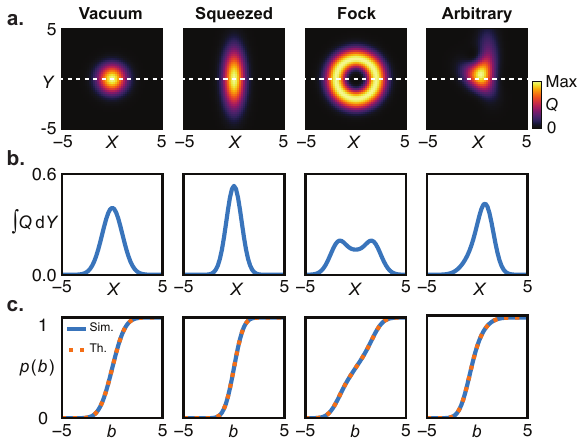}
    \caption{\small \textbf{Shaping bias-probability relationship with quantum statistics.} \textbf{a.} Husimi $Q$-function of the initial state. Dashed line corresponds to $Y=0$. \textbf{b.} Corresponding marginal $Q$-function (integrated along $Y$ quadrature). \textbf{c.} Corresponding bias-probability relationship of the steady state. }    
    \label{fig:2}
    \vspace{-0.3cm}
\end{figure}

\begin{figure}
\centering
\vspace{-0.2cm}
  \includegraphics[scale=0.9]{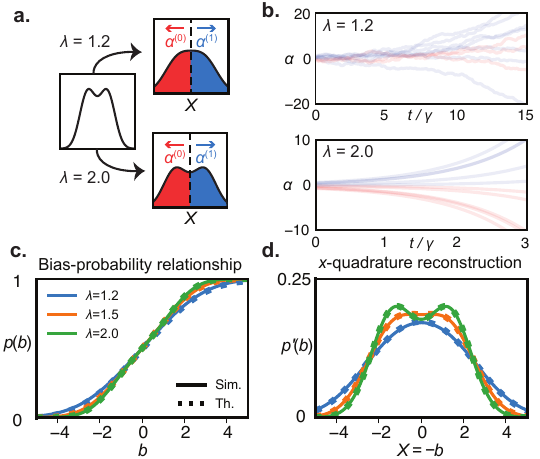}
    \caption{\small \textbf{Influence of quantum noise on steady-state probabilities.} \textbf{a.} Quantum statistics of the initial state can be washed out depending on the value of the noise ($\lambda = 2$ corresponds to the noiseless regime). \textbf{b.} Corresponding examples of early-time stochastic trajectories. The color of the line corresponds to the sign of the initial value of the quadrature (- red, + blue). \textbf{c.} Resulting bias-probability relationship for different values of noise. The initial state is a cat-state $|\alpha\rangle + |-\alpha\rangle$ with amplitude $\alpha = 1$. \textbf{d.} Reconstructed $X$-quadrature of the Husimi $Q$-function.}
    \label{fig:noise}
    \vspace{-0.3cm}
\end{figure}


By integrating the stochastic differential equation for the $X$ quadrature, we obtain an analytical expression for the bias-probability relationship under a quantum initial condition:
\begin{equation}\label{steadystate}
p(\alpha^{(1)}, b) = \int_{-\frac{\sqrt{2}b}{\lambda-1}}^\infty p_{X_0} * g_\sigma(x) ~\text{d}x.
\end{equation}
Here, $p_{X_0}$ denotes the $X$-marginal of the phase space distribution of the initial quantum state, $*$ denotes convolution, and $g_\sigma$ is a normalized Gaussian with variance $\sigma^2 = (1 + \xi(1-\lambda))/(2(\lambda-1))$. This result shows that the steady state distribution $p(\alpha^{(1)})$ is obtained by taking the initial $X$-quadrature marginal distribution, smoothing by a Gaussian of variance $\sigma^2$, and then taking the right half of the resulting distribution.

This interplay between the initial condition's quantum statistics and steady-state probability is directly reflected in the bias-probability relationship, shown for different initial quantum states in Fig.~\ref{fig:2}. Varying the amplitude of the coherent bias $b$ in the initial quantum state corresponds to shifting the effective ``decision boundary'' of the steady-state probabilities in phase space. Applying a coherent bias displaces the state in phase space, so $p(\alpha^{(0)})$ now corresponds to the area of the portion of the marginal $X$-distribution that lies to the right of this shifted boundary. As a result, the shape of the bias-probability relationship mirrors the cumulative distribution function of the marginal $X$-distribution of the initial quantum state in phase space (Fig.~\ref{fig:2}(c)). This is the key observation which generally allows us to understand the quantum sensitivity of parametric oscillation.

Simulations of the stochastic trajectories of the $X$ quadrature governed by Eq.~(\ref{eq:sde-maintext}) offer further insight into the bias-probability relationship (Fig.~\ref{fig:noise}(b)). 
The phase-sensitive gain causes the quadrature trajectories to grow in time, as is expected from parametric amplification. Additionally, both the gain and dissipation can add noise to the field, resulting in fluctuations from a purely exponential behavior. This effect is evident for $\lambda = 1.2$, where we observe ``crossing over'' of trajectories between the two phases (trajectories with initial $X>0$ may end up at $\alpha^{(0)}$, and vice versa). The noise effectively washes out the shape of the marginal $X$-distribution of the phase-space function as the initial configuration propagates into the steady state distribution, as depicted in Fig.~\ref{fig:noise}(a). For the $Q$-function, the semi-classical stochastic trajectories has a noise coefficient which vanishes at $\lambda = 2$, and the $Q$-function trajectories evolve deterministically based on their initial $X$-value. In other words, trajectories that start with $X>0$ end up at $\alpha^{(1)}$, while trajectories starting at $X < 0$ end up at $\alpha^{(0)}$: the system's (fluctuating) initial condition fully determines its steady state. In the $Q$-function, this means that all of the quasiprobability mass that starts on the right ($X > 0$) arrives at the steady state $\alpha^{(1)}$, and similarly for the other steady-state. We note that, although the trajectories of the Husimi $Q$ representation appear to follow deterministic paths, this function describes a particular coherent state representation and thus the density matrix it represents still experiences quantum noise.

In general, the initial phase-space distribution will be Gaussian-filtered with a filter width $\sigma$, due to the Langevin noise induced by the cavity decay. This relationship is depicted in Fig.~\ref{fig:noise}(c,d), where we simulate the bias-probability relationship $p(b)$ for an initial cat state at various values of $\lambda$. Based on the above theory, the derivative of $p(b)$ should correspond to a smoothed version of the initial $X$-quadrature. 
We simulated this reconstruction process for various other representative initial states in Fig.~\ref{fig:2}(a-c), where we find that in each case, the bias-probability relationship corresponds to the marginal $X$-distribution (of the $Q$ function). For example, for the Gaussian states (vacuum and squeezed), the bias-probability relationship is an error-function, as expected. In each case, it is clear how the properties of the initial quantum state are reflected in the steady-state statistics that we observe macroscopically. In principle, we could also rotate the pump phase and repeat this procedure across various phases---thereby reconstructing the entire initial $Q$-function, which provides a complete description of the quantum state~\cite{choi2024intracavity}.

\subsection*{Quantum initializations of Josephson parametric oscillators}

\begin{figure}
\centering
\vspace{-0.2cm}
  \includegraphics[scale=0.7]{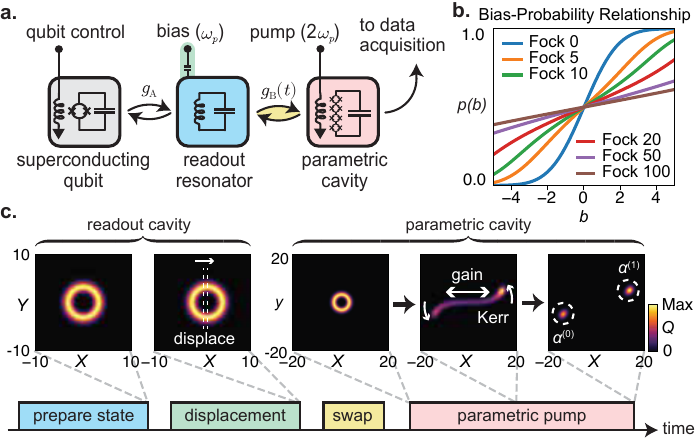}
    \caption{\small \textbf{Experimental proposal with superconducting qubits.} \textbf{a.} Schematic of the proposed experimental setup in superconducting qubits. \textbf{b.} Bias-probability relationship for different photon number Fock states injected into the parametric cavity. \textbf{c.} Experimental protocol for 5-photon Fock state biasing.}
    \label{fig:3}
    \vspace{-0.3cm}
\end{figure}

Finally, we describe how this type of behavior is not limited to OPO systems at optical frequencies, by showing the same type of response in superconducting microwave circuits. Superconducting circuits support strong nonlinearities enabled by the Josephson effect, which can be used to generate and manipulate complex quantum states such as high-photon Fock states, cat states, and other non-classical states~\cite{fink2008climbing,hofheinz2008generation,hofheinz2009synthesizing, eichler2011experimental, vlastakis2013deterministically}, and which are a leading platform for quantum information processing \cite{devoret2013superconducting}. As such, superconducting circuits are well suited for non-Gaussian quantum state generation beyond squeezed states, and are also well suited for measurement of these states using quantum-limited parametric amplifiers \cite{aumentado2020superconducting}.

In Fig.~\ref{fig:3} we consider a microwave resonant circuit that contains one or more superconducting quantum interference devices (SQUIDs, which are Josephson junction-based circuit elements) that act as both a Kerr nonlinearity, and, a linear inductance that can be dynamically modulated by magnetic flux. When flux-modulated near twice the resonator's natural frequency, the resulting circuit can be described by the Heisenberg-Langevin equations of motion \cite{wustmann2013parametric,wustmann2019parametric}:
\begin{equation}\label{kerr}\dot{a} = -a + \lambda a^\dagger + ig^2(a^\dagger a)a + F(t) +b,\end{equation}
where $\lambda$, $g$, $F(t)$, and $b$ have the same definitions as in Eq.~(\ref{eq:1}).
Like in an OPO, this system experiences two steady states when pumped above threshold, due to the intensity dependent detuning imposed by the Josephson-junction based Kerr nonlinearity, which functions in a similar way as the $\chi^{(2)}$ nonlinearity in an OPO. Such systems have thus been dubbed JPOs (Josephson parametric oscillators) \cite{krantz2013investigation} and may be used for superconducting qubit readout \cite{lin2014josephson,krantz2016singleshot,rosenthal2021efficient}. Like an OPO, the JPO has high sensitivity to the initial quantum state, except that the steady state distribution is rotated in phase space due to its Kerr nonlinearity (Fig~\ref{fig:3}(c)). 

Fig.~\ref{fig:3}(a) shows one way of setting up a superconducting microwave version of this experiment, making use of the device demonstrated in Ref.~\cite{rosenthal2021efficient}. First, a quantum state is prepared in a readout resonator using standard techniques from circuit quantum dynamics. Next, this state is then swapped into a parametric cavity using a time-dependent coupling $g_B(t)$. Finally, the parametric cavity is pumped to amplify its field and measure its steady state statistics. For this setup, Fig.~\ref{fig:3}(b) shows simulations of the bias-probability relationship for various Fock states, which agrees well with our analytical model (details in SI, Section S7). Finally, note that alternative versions of this experiment could instead make use of parametric amplifiers directly integrated into qubit readout cavities \cite{eddins2019highefficiency}.

\section*{Discussion}
We have analyzed the sensitivity of parametric oscillators to quantum initial conditions and have found that they generally exhibit a high sensitivity to initial conditions that allows the imprinting of initial conditions' quantum statistics into a macroscopic steady-state observable. By measuring the sensitivity to an externally injected coherent bias field, we are able to trace back to the system's initial quantum signature. Generally, we expect our results to generalize to multistable driven-dissipative systems in the low quantum noise regime, where early-stage linearized dynamics determine steady-state statistics. 

We have proposed two platforms for the observation of these effects, namely free-space or integrated OPOs, and a Josephson parametric amplifier. Given the current capabilities of both platforms, we envision that the latter may result in a shorter term experimental demonstration, given the relative maturity of superconducting qubit technologies in (1) generating non-Gaussian quantum states~\cite{vlastakis2013deterministically, fink2008climbing}; (2) high-fidelity swapping and measurement protocols of quantum states between cavities~\cite{rosenthal2021efficient, steffen2006measurement}; (3) availability of bistable nonlinear parametric amplifier based on the Josephson effect~\cite{aumentado2020superconducting}. 

Given the recent interest in driving light-matter systems with quantum light~\cite{leroux2018enhancing, gorlach2023high, heimerl2024multiphoton} (mostly restricted to (bright) squeezed states of light), one could expect that, more generally, quantum statistics of the drive may be imprinted on the behavior of these systems. 
We also expect that generalizations of this theory to multimodal systems, such as Kerr combs and synchronously pumped OPOs~\cite{cai2017multimode, guidry2023multimode, lustig2024emerging, pontula2024multimode}, where multimode quantum correlations may appear, will lead to greater complexity in mapping quantum statistics to classical observables, and may also enable tomography of multimode entangled states. Our work introduces a new class of experiments that can test how sensitive macroscopic systems are to quantum initial conditions and provides a method for controlling systems with quantum degrees of freedom.


\section*{Competing interests}
The authors declare no potential competing financial interests.

\section*{Data and code availability statement}
The data and codes that support the plots within this paper and other findings of this study are available from the corresponding authors upon reasonable request. Correspondence and requests for materials should be addressed to A.~G. (alexgu@mit.edu) and C.~R.-C. (chrc@stanford.edu).

\section*{Acknowledgments}

We acknowledge useful discussions with Melissa Guidry and Eran Lustig. J.S. acknowledges previous support of a Mathworks Fellowship, as well as previous support from a National Defense Science and Engineering Graduate (NDSEG) Fellowship (F-1730184536). C.R.-C. is supported by a Stanford Science Fellowship. E. I. R. acknowledges support by an appointment to the Intelligence Community Postdoctoral Research Fellowship Program at Stanford University administered by Oak Ridge Institute for Science and Education (ORISE) through an interagency agreement between the U.S. Department of Energy and the Office of the Director of National Intelligence (ODNI). M.~H. acknowledges funding from the Austrian Science Fund (FWF) through grant J4729. This material is based upon work sponsored in part by the U.S. Army DEVCOM ARL Army Research Office through the MIT Institute for Soldier Nanotechnologies under Cooperative Agreement number W911NF-23-2-0121 and by the Vannevar Bush Faculty Fellowship from the Department of Defense. We acknowledge the MIT SuperCloud and Lincoln Laboratory Supercomputing Center for providing (HPC, database, consultation) resources that have contributed to the research results reported within this paper/report.

\bibliographystyle{ieeetr}
\bibliography{main.bib}

\end{document}